%% file: main.tex
\documentclass[a4paper]{article}

\usepackage{INTERSPEECH2021}
\usepackage{url}
\usepackage{xcolor}
\usepackage{cite}
\usepackage{float}
\usepackage{booktabs,multirow}
\usepackage{pifont}
\newcommand{\cmark}{\ding{51}}%
\usepackage{balance}
\usepackage{hyperref}
\usepackage{amsmath,graphicx}

\title{Muskits: an End-to-End Music Processing Toolkit for Singing Voice Synthesis}
\name{Jiatong Shi$^1$, Shuai Guo$^2$, Tao Qian$^2$, Nan Huo$^3$, Tomoki Hayashi$^{4,5}$, Yuning Wu$^2$, Frank Xu$^1$, Xuankai Chang$^1$, Huazhe Li$^6$, Peter Wu$^7$, Shinji Watanabe$^1$, Qin Jin$^2$$^{\dagger}$\thanks{\scriptsize{$^\dagger$Corresponding Author.}}}
\address{
  $^1$Carnegie Mellon University, $^2$Renmin University of China, $^3$The University of Hong Kong, \\ $^4$Human Dataware Lab. Co., Ltd., $^5$Nagoya University, $^6$Tsinghua University, \\ $^7$University of California, Berkeley}
\email{jiatongs@cs.cmu.edu, \{shuaiguo, qiantao, qjin\}@ruc.edu.cn, shinjiw@ieee.org}

\begin{document}

\maketitle
\input{sections/0_Abstract}

\input{sections/1_Introduction}

\input{sections/2_Functionality}

\input{sections/3_Experiments}

\input{sections/4_Conclusion}


\bibliographystyle{IEEEtran}

\bibliography{mybib}


\end{document}

%% file: sections/0_Abstract.tex
\begin{abstract}
\vspace{-5pt}
  This paper introduces a new open-source platform named \textbf{\textit{Muskits}} for end-to-end music processing, which mainly focuses on end-to-end singing voice synthesis (E2E-SVS). Muskits supports state-of-the-art SVS models, including RNN SVS, transformer SVS, and XiaoiceSing. The design of Muskits follows the style of widely-used speech processing toolkits, ESPnet and Kaldi, for data prepossessing, training, and recipe pipelines. To the best of our knowledge, this toolkit is the first platform that allows a fair and highly-reproducible comparison between several published works in SVS. In addition, we also demonstrate several advanced usages based on the  toolkit functionalities, including multilingual training and transfer learning. This paper describes the major framework of Muskits, its  functionalities, and experimental results in single-singer, multi-singer, multilingual, and transfer learning scenarios. The toolkit is publicly available at \url{https://github.com/SJTMusicTeam/Muskits}.
\end{abstract}
\vspace{-15pt}

%% file: sections/1_Introduction.tex
\section{Introduction}

Singing voice synthesis (SVS) uses music score and lyrics to generate natural singing voices of a target singer. Recently, SVS has attracted much research attention in the speech and music processing communities. However, due to high data annotation costs and strict musical copyright policies, there are limited public databases available for SVS. The shortage of high-quality data further leads to limited benchmarks in SVS, compared to various open-source activities in other generative tasks such as text-to-speech (TTS) and voice conversion (VC) \cite{schroder2011open, hayashi2021espnet2, tts_mozilla, tts_coqui}.

This paper introduces a new open-source toolkit called \textit{\textbf{Muskits}} that aims to provide a neural network-based end-to-end platform for music processing, specifically SVS. Unlike existing open-source SVS repositories \cite{oura2010recent, nnsvs, chandna2019wgansing, tae2021mlp, choi2020korean, shi2021sequence, wang2022opencpop}, Muskits provides different architectures for SVS in an end-to-end manner and offers a wide-range comparison between different architectures for singing voice synthesis. Muskits inherits the base framework from ESPnet \cite{hayashi2021espnet2, watanabe2018espnet} and follows ESPnet styles in data processing scripts and training recipes, demonstrating a complete setup for SVS. The contributions of this paper include: (1) a new platform for reproducible SVS modeling; (2) benchmarks on several public databases in both single- and multi-singer scenarios; (3) exploration on multilingual training for SVS; (4) studies on transfer learning 
on a multi-style low-resource database.

\section{Related Works}
The initial effort on SVS starts with unit-selection models, where pre-recorded singing segments are concatenated to form the singing voice \cite{kenmochi2007vocaloid, bonada2016expressive}. However, because of the need for large corpora, this method is not flexible to achieve reasonable sound in low-resource scenarios. Later, statistical parametric methods (e.g., the hidden Markov model (HMM)), were proposed for singing synthesis \cite{saino2006hmm, oura2010recent}. They were shown to flexibly synthesize the singing voice, but also led to loss of naturalness. Deep neural networks (DNNs) have become dominant across several generation tasks in the recent decade. SVS is not an exception. Initially, several works applied basic structures, including fully-connected layers, convolutional neural networks (CNNs), and recurrent neural networks (RNNs) \cite{nishimura2016singing, kim2018korean, nakamura2019singing, hono2018recent}, which demonstrated reasonable quality and naturalness for singing generation. Some more advanced architectures, including generative adversarial networks (GANs), encoder-decoder, and diffusion-based models, also showed some quality improvements over specific databases \cite{blaauw2020sequence, gu2021bytesing, lu2020xiaoicesing, hono2019singing, lee2019adversarially, liu2021diffsinger, zhang2021visinger}.

As SVS is receiving more and more attention, there has been an increasing need for open-source platforms so as to allow fair comparison between the proposed models. Compared to similar tasks such as TTS, SVS has faced several issues in sharing benchmarks and their platforms. Different from speech, singing voice databases usually have more strict usage guidelines and copyright concerns. Meanwhile, singing voices require a professional high-quality recording environment and generally take longer to record than that for regular speech. Resources are further limited due to the need for detailed annotation (e.g., music score information).

Even with those limitations, there are still some relevant efforts in releasing databases and code bases for some SVS models. Sinsy \cite{oura2010recent, hono2018recent}, known as one of the initial open-source works in SVS, was released for research purposes. The toolkit applies the HMM-based architecture and can reach reasonable performances with low-resource singing voice corpus (e.g., 1 hour). It also demonstrates its extendability to multiple languages, including Japanese, Chinese, and English. NN-SVS \cite{nnsvs} is a DNN-based singing voice synthesis toolkit, which follows the cascaded neural networks in \cite{hono2018recent}. Seven Japanese recipes are supported in the toolkit. We also observe several open-source efforts related to specific models, including \cite{chandna2019wgansing, tae2021mlp, choi2020korean, shi2021sequence, wang2022opencpop}. They mostly include the whole training procedure and some also release the related databases \cite{tae2021mlp, wang2022opencpop}. 

The aforementioned open-source efforts have greatly benefited the whole research community. However, it is difficult to provide a fair comparison of different models due to the differences in datasets, feature representations, and vocoders. In this work, our goal is to fill the gap with a new open-source toolkit, Muskits. With 12 recipes on 10 SVS databases, the toolkit has supported several choices of end-to-end SVS models with various ESPnet compatible vocoders\footnote{\scriptsize {\url{https://github.com/kan-bayashi/ParallelWaveGAN}}}. Furthermore, we explore multilingual training, pre-training, and transfer learning on a multi-style databases using Muskits, which have not been well investigated in SVS before.

%% file: sections/2_Functionality.tex
\vspace{-10pt}
\section{Functionality}
Muskits consists of two major parts: a Python library of several neural network models and recipes for running complete experiments on specific databases. The library uses PyTorch for neural networks, with standard music representation, various model architectures, and a uniform training/decoding framework. The recipes provide detailed examples in all-in-one style scripts, following the style in Kaldi and ESPnet \cite{hayashi2021espnet2, watanabe2018espnet, povey2011kaldi}. 

\subsection{Music score representation}
\label{ssec: music representation}
Muskits performs all pre-processing steps on the fly, including acoustic feature extraction and two standard interfaces for music score information. The two interfaces are at the frame-level and syllable-level. The frame-level features need pre-aligned linguistic units (e.g., phoneme or syllable) and music note sequences. Based on timestamps, both sequences are first expanded to the sampling rate of a given audio (e.g., 24k). Then, they are aggregated with a sliding window that has the same window size and hop length to the target acoustic features (e.g., log-Mel filter-bank features). The same or similar features are used in \cite{shi2021sequence, kim2018korean, nakamura2019singing, hono2018recent, tae2021mlp}. The syllable-level features are an extended version of XiaoiceSing feature \cite{lu2020xiaoicesing}, including phoneme ids, music note ids\footnote{The music note ids are the same as MIDI 0-128 note ids.}, tempo/beats information, and corresponding duration information. Depending on the models and application scenarios, either frame-level or syllable-level features could be selected. Note that we always use syllable-level features for models that consider duration modeling (e.g., various sequence-to-sequence models) \cite{chen2020hifisinger, blaauw2020sequence, lu2020xiaoicesing, shi2021sequence}.

\subsection{Models}
Muskits supports three models, including a RNN-based model \cite{kim2018korean, shi2021sequence}, transformer-based model \cite{blaauw2020sequence}, and XiaoiceSing \cite{lu2020xiaoicesing}. All models are based on the encoder-decoder architecture and support both frame and syllable-level representations discussed in Sec~\ref{ssec: music representation}. The output of models is a sequence of acoustic features (e.g., Mel-filter bank features).

Similar to \cite{lu2020xiaoicesing, chen2020hifisinger, shi2021sequence}, in the RNN-based model, music notes and phoneme information are first passed to individual RNN encoders. Then, the combination of hidden states (either by concatenating or adding) is fed into another RNN decoder. If syllable-level representations are used for the RNN-based model, an additional duration predictor is introduced to expand the syllable-level hidden states to the frame level. The duration predictor is trained separately to predict the explicit duration of each hidden state. In training, the encoder-decoder model adopts ground truth duration. While in inference, the encoder output will be expanded by the predicted duration, and then fed into the decoder. The transformer-based model follows the architecture of \cite{blaauw2020sequence}. It first encodes the phoneme sequence by stacked gated linear unit (GLU) layers. Then, it expands the hidden representation of phonemes into frame-level with a rule-based duration model. After adding the pitch embedding and positional encoding, the hidden states are then fed into a transformer-based decoder for final outputs. The duration of phoneme sequences is defined using rules \cite{blaauw2020sequence}. XiaoiceSing adopts a FastSpeech-like architecture \cite{lu2020xiaoicesing, ren2020fastspeech}. The model sums over the syllable feature representations and adopts an encoder-decoder framework for singing voice generation.

\subsection{Training, inference, and evaluation}
\label{ssec: train, inference, eval}
The training procedures in Muskits are handled with a unified task processor adapted from ESPnet, which supports multi-GPU training and dynamic batch making. On-the-fly data augmentation is supported, including pitch augmentation and mixup-augmentation \cite{guo2022singaug}.

In addition to the default Griffin-Lim vocoder, Muskits also supports various neural vocoders, including ParallelWaveGAN, MelGAN, and HiFiGAN \cite{yamamoto2020parallel, kumar2019melgan, yang2021multi, mustafa2021stylemelgan, kong2020hifi}. The vocoder can be applied to both training and inference stages. In training, the vocoder could automatically generate audio from selected samples of development set for monitoring the training progress. 
During inference, compatible vocoders can be used directly without additionally dumping intermediate features.

We implement three objective evaluation metrics for SVS in Muskits: Mel-cepstral distortion (MCD), voiced/unvoiced error rate (VUV\_E), and logarithmic rooted mean square error of the fundamental frequency (F$_0$RMSE). All the metrics are calculated by dynamic time-warping to consider the length mismatch between synthesized singing and ground truth singing.

\subsection{Recipe flow}
We provide 12 all-in-one recipes for 10 singing voice databases. All recipes follow a unified pipeline with explicit stages of preparation, training, inference, and evaluation. The stages are defined in the template \texttt{svs.sh}, as follows:

\noindent
\textit{\textbf{Stage 1}: Database-dependent data preparation}. The expected data generally follows Kaldi style but with additional \texttt{midi.scp} and \texttt{label} for music score and phoneme alignment information. For databases without explicit midi or MusicXML, we also provide rule-based automatic music transcription to extract related music information. Relevant functions can be found in KiSing recipe \cite{shi2020kising}.

\noindent
\textit{\textbf{Stage 2}: Standard audio and midi formatting}. This stage formats audio into a standard format by resampling, segmentation, and dumping from pipe-style input. MIDI is also normalized and segmented at this stage. Singer ID and Language ID are also defined if needed.

\noindent
\textit{\textbf{Stage 3}: Filtering}. Phrases in the training and development sets are filtered by a pre-defined length threshold.

\noindent
\textit{\textbf{Stage 4}: Token list generation}. Muskits collects tokens from the training set and generates a corresponding token list for training.

\noindent
\textit{\textbf{Stage 5}: Statistics collection}. The input information of training and development sets are collected for efficient batching.

\noindent
\textit{\textbf{Stage 6}: Model training}. The model is optimized based on the training objectives. 

\noindent
\textit{\textbf{Stage 7}: Model inference}. If no vocoder is provided, Muskits adopts the Griffin-Lim vocoder. Otherwise, compatible vocoders can be loaded for inference. The predicted acoustic features are also stored for external vocoders.

\noindent
\textit{\textbf{Stage 8}: Objective evaluation}. Objective evaluation is conducted between paired reference data and inference outputs.

\noindent
\textit{\textbf{Stage 9}: Model packing}. The model is packed with all the necessary elements for inferences. The packed models of our recipes are also shared with the recipes.

Most public SVS databases only have a single singer, which limits the scalability of the SVS systems built upon them. By supporting several recipes with a standardized training interface, Muskits could easily extend to different genders, multi-singer, multilingual scenarios. To be specific, we already support two recipes on multi-singer and multilingual cases.

%% file: sections/3_Experiments.tex
\vspace{-10pt}
\section{Experiments}

\subsection{Experimental datasets}
\label{ssec: dataset}
To demonstrate the performance of Muskits, we conduct four sets of experiments under single-singer, multi-singer, multilingual, and transfer learning scenarios. For single-singer experiments, we use the Ofuton-P database \cite{futon2021ofuton} that contains 56 Japanese songs (61 minutes) of a male singer. For multi-singer experiments, we use a combination of four databases, including Ofuton-P, Oniku, Natsume, and Kiritan \cite{futon2021ofuton, kurumi2021oniku, amanokei2020natsume, moris2020kitian}. For multilingual training, we adopt the four databases mentioned above, CSD databases \cite{choi2020children}, and Opencpop \cite{wang2022opencpop}, which results in a ten-hour corpus in Japanese, English, Korean, and Mandarin. For transfer learning, we adopt the KiSing corpus, a one-hour female Mandarin corpus with multiple styles \cite{shi2020kising}.

Except for Opencpop and KiSing, other databases do not have official splits of train, development, and test sets, so we apply our own split in our reproducible recipes. All the songs are segmented based on the silence between lyrics. The data sizes for each experiment are listed in Table~\ref{tab: dataset}. For CSD, only syllables are provided for the alignment, but the training set is not enough to cover all the syllable combinations. Therefore, we split the syllable into phonemes and evenly assign duration information to each phoneme given the duration of syllables.

\begin{table}
\centering
\caption{\label{tab: dataset} Details of experimental data: \textbf{N$_{\text{src}}$}, \textbf{N$_{\text{lang}}$}, \textbf{N$_{\text{singer}}$} stands for number of databases, languages, and singers, respectively.}
\vspace{-6pt}
\begin{tabular}{l|c|ccc}
\toprule \multirow{2}{*}{\textbf{Task}} & \multirow{2}{*}{\textbf{N$_{\text{src}}$, N$_{\text{lang}}$, N$_{\text{singer}}$}} & \multicolumn{3}{|c}{\textbf{Duration(h)}} \\
&  & \textbf{Train} & \textbf{Dev} & \textbf{Test} \\ 
\midrule
Single-singer & 1,1,1 & 0.63 & 0.08 & 0.07 \\
Multi-singer  & 4,1,4 & 3.01 & 0.34 & 0.38 \\
Multilingual  & 6,4,6 & 11.72 & 0.49 & 1.01\\
Transfer      & 1,1,1 & 0.71 & 0.02 & 0.05\\
\bottomrule
\end{tabular}

\vspace{-15pt}
\end{table}

\vspace{-4pt}
\subsection{Experimental setups}
\label{ssec: exp_setup}
\vspace{-4pt}
The experiments are in two folds based on whether the model uses ground truth (G.T.) duration in order to factorize the evaluation of duration prediction and acoustic modeling. 

The RNN-based model adopts three-layer 512-dimension bi-directional long short time memory (LSTM) encoder and a five-layer 1024-dimension decoder. The transformer-based model follows the original setting in \cite{blaauw2020sequence}. It consists of three 256-channel GLU layers with a kernel size of three as the encoder and a six-layer four-head-256-dimension transformer decoder. Local Gaussian constraints are injected into the self-attention in its decoder. The XiaoiceSing adopts the same network hyper-parameters as in \cite{lu2020xiaoicesing}, including an encoder and a decoder with the same setting (i.e., six transformer blocks, while each block has four-head-384-dimension self-attention block and 1536 dimensional feed-forward layer). 
The decoder outputs of all models are projected to the dimension of acoustic features (i.e., 80) with a postnet inherited from Tacotron2 \cite{shen2018natural}. For settings with duration prediction, the transformer-based model adopts a rule-based duration modeling as defined in \cite{blaauw2020sequence}, while the others adopt two 384-channel convolutional layers with a kernel size of three to model duration information. 

For multi-singer and multilingual training, we provide the option to apply additional embeddings from one-hot singer IDs and language IDs. The embedding size is the same as the dimension of the encoder outputs for all models. In the forward process, embeddings are added directly to the hidden states of encoders. As discussed in Sec~\ref{ssec: train, inference, eval}, the acoustic models can be easily used with compatible vocoders. We used the HiFi-GAN vocoder for all experiments. The vocoders are trained on the same training set from the Muskits' split introduced in Sec~\ref{ssec: dataset}.

During training, the RNN-based model uses Adam optimizer with a 0.001 learning rate without learning rate schedulers to keep the same as its original paper \cite{shi2021sequence}. All the other models employ Adam optimizer with a NoamLR scheduler. The warm-up steps are set to 4,000, and the peak learning rate is set to 1.0. We employ gradient clipping at the threshold of 1.0. All experiments are conducted on a single GPU and the batch size is set to 8. We apply a 0.1 dropout rate to encoder/decoder and a 0.5 dropout rate to postnets. We set the training epochs as 500 for models trained on single-singer databases and 250 for models trained on multi-singer or multilingual data.

As in Sec~\ref{ssec: train, inference, eval}, we adopt MCD, V/UV Error, and F$_0$RMSE for objective evaluation. For each subjective evaluation conducted online, we invite 25 subjects, including both musicians and non-professionals. The same set of 15 phrases are randomly selected for each system. Subjects are asked to score the phrases in one (unintelligible) to five (excellent quality). We report our results with 95\% confidence intervals. We also conduct the A/B test in the transfer learning scenario. The same 25 subjects are asked to compare two samples of the same phrase from a pair of models. 20 phrases are selected from the KiSing test set.

\subsection{Results and discussion}

\noindent\textbf{{Single singer:}}
Table~\ref{tab: models exp} presents the results of the single-singer setting. Among models using G.T. duration, the RNN-based model achieves the best performance over the subjective MOS metric; the Transformer-based model achieves the best MCD; XiaoiceSing achieves the best VUV\_E and F$_0$RMSE. For models without G.T. duration, the RNN-based model achieves the best performance on both objective and subjective metrics. We observe that there is a huge gap between models with and without G.T. duration. The primary reason could be the limited size of the Ofuton-P database (37.8 minutes of singing for training), which causes the difficulty of learning  reasonable duration predictors. Moreover, the results in the subjective evaluation suggest that the RNN-based model is better than the other two, and the Transformer-based model is better than XiaoiceSing, in our single-singer scenario. This may indicate that simpler structures are more suitable for the task with low data-resource.

\begin{table*}
\centering
\caption{\label{tab: models exp}: Single- and multi-singer experiments: The experimental databases are Ofuton-P database for single-singer experiments and a combination of four databases as described in Sec~\ref{ssec: dataset}; the models are discussed in Sec~\ref{ssec: exp_setup}; the metrics are defined in Sec~\ref{ssec: train, inference, eval}.  }
\vspace{-6pt}
\begin{tabular}{l|c|cccc|cccc}
\toprule \multirow{2}{*}{\textbf{Model}} & \textbf{G.T.} & \multicolumn{4}{c|}{\textbf{Single-Singer}} & \multicolumn{4}{c}{\textbf{Multi-Singer}} \\

&  \textbf{Dur.} & \textbf{MCD$\downarrow$} & \textbf{VUV\_E$\downarrow$} & \textbf{F$_0$RMSE$\downarrow$} & \textbf{MOS$\uparrow$} & \textbf{MCD$\downarrow$} & \textbf{VUV\_E$\downarrow$} & \textbf{F$_0$RMSE$\downarrow$} & \textbf{MOS$\uparrow$}\\ 
\midrule
RNN \cite{shi2021sequence}  & \cmark & 6.52 & 2.30 & 0.105 & \textbf{3.47 $\pm$ 0.11} & 7.15 & 3.87 & 0.195 & 3.25 $\pm$ 0.13 \\
Transformer \cite{blaauw2020sequence} & \cmark & \textbf{6.46} & 2.67 & 0.118 & 3.23 $\pm$ 0.11 & \textbf{6.68} & 3.90 & \textbf{0.184} & \textbf{3.40 $\pm$ 0.12} \\
XiaoiceSing \cite{lu2020xiaoicesing} &\cmark & 6.88 & \textbf{2.22} & \textbf{0.103} & 3.11 $\pm$ 0.11 & 6.81 & \textbf{3.86} & 0.190 & 3.23 $\pm$ 0.12\\
\midrule
RNN \cite{shi2021sequence} & / & \textbf{6.19} & \textbf{2.18} & \textbf{0.109} & \textbf{3.43 $\pm$ 0.11} & 7.05 & \textbf{3.55} & 0.226 & 2.84 $\pm$ 0.12 \\
Transformer \cite{blaauw2020sequence}  & /  & 6.95 & 5.61 & 0.172 & 2.39 $\pm$ 0.11 & 6.87 & 3.70 & 0.186 & \textbf{3.33 $\pm$ 0.12} \\
XiaoiceSing \cite{lu2020xiaoicesing}  & /  & 7.19 & 2.68 & 0.120 & 1.96 $\pm$ 0.08  & \textbf{6.79} & 3.85 & \textbf{0.154} & 2.83 $\pm$ 0.11\\
\midrule 
G.T. & / & / & / & / & 4.89 $\pm$ 0.05 & / & / & / & 4.65 $\pm$ 0.10\\
\bottomrule
\end{tabular}

\vspace{-10pt}
\end{table*}

\vspace{3pt}
\noindent\textbf{{Multiple singer:}}
Table~\ref{tab: models exp} presents the experiment results of the multi-singer experiments setting. Even though multi-singer corpus has larger data size compared to single-singer, it may encounter additional issues like multiple styles of singing and singer variation. For example, we find singing voices in Oniku database mostly use falsetto, while the other databases use mostly modal voice. As in Table~\ref{tab: models exp}, evaluation metrics on multi-singer may not be necessarily better than single-singer's. However, as we compare the MOS between single and multi-singer scenarios\footnote{Other metrics are not suitable for comparison, given that the test data is different.}, we observe that the RNN-based model has some degradation on MOS, while the other two have some improvements. It might indicate that the transformer-based model and XiaoiceSing have a better modeling capacity and need a certain amount of data to converge. The RNN-based model, on the other hand, can efficiently converge on small data, but has the capacity issue when extending to multi-singer scenarios.


\vspace{3pt}
\noindent\textbf{{Multilingual singing:}}
We conduct five experiments on multilingual SVS and evaluate the performances on Ofuton-P database. The baseline is the XiaoiceSing without duration prediction presented in Table~\ref{tab: models exp}. The pre-train models are trained on the multilingual data introduced in Sec~\ref{ssec: dataset} with the same XiaoiceSing architecture. As in Sec~\ref{ssec: exp_setup}, we examine models with or without language IDs. Based on models trained on multilingual data, we also fine-tune the multilingual models on Ofuton-P database for another 500 epochs.  As in Table~\ref{tab: multilingual}, the multilingual models clearly demonstrate a large gain over the MCD objective scores, but do not improve the V/UV error rate and F$_0$RMSE. 
In subjective evaluation, directly using multilingual models does not bring improvements on MOS. However, pre-trained models could be useful when applying fine-tuning on the same database. Meanwhile, given that the model with language IDs always getting worse objective and subjective scores, it might indicate that additional language IDs could not improve the singing quality. One potential reason is that language IDs may restrict the model to transfer knowledge from multilingual data especially when the data size is in low-resource.

\begin{table}
\centering
\caption{\label{tab: multilingual} Multilingual experiments on Ofuton-P. All the models use the same XiaoiceSing architecture. 
The pre-trained models are conducted over the multilingual data described in Sec.~\ref{ssec: dataset}.
}
\vspace{-10pt}

\begin{tabular}{l|cccc}
\toprule \textbf{Model} & \textbf{MCD$\downarrow$} & \textbf{VUV\_E$\downarrow$} & \textbf{F$_0$RMSE$\downarrow$} & \textbf{MOS$\uparrow$} \\ 
\midrule
Baseline    & 6.88 & \textbf{2.22} & \textbf{0.103} & 3.05 $\pm$ 0.12 \\
Pre-train   & \textbf{6.17} & 2.51 & 0.113 & 2.99 $\pm$ 0.12\\
\  + LangID & 6.32 & 2.94 & 0.123 & 2.90 $\pm$ 0.12\\
Fine-tune   & 6.23 & 2.75 & 0.106 & \textbf{3.49 $\pm$ 0.12} \\
\  + LangID & 6.34 &2.55 & 0.110 & 3.09 $\pm$ 0.12\\
\midrule
G.T. & / & / & / & 4.77 $\pm$ 0.07 \\
\bottomrule
\end{tabular}
\vspace{-15pt}
\end{table}

\vspace{3pt}
\noindent\textbf{{Transfer learning:}}
In this experiment, we explore to use transfer learning on KiSing, a low-resource multi-style database. Although KiSing has 42.6 minutes of singing for training, it has content duplication in most songs, resulting in limited variances in terms of phoneme sequences and melodies. Furthermore, the singer has adopted both a Chinese folk style (a variant of the Peking opera) and a pop style in songs, which has more frequent changes in rhythm, harmony, and the usage of falsetto. We conducted three experiments on this database: Baseline is trained only on KiSing. Two pre-trained models Multilingual$*$ and Opencpop$*$ are first pre-trained on multilingual data and Opencpop data, respectively. And then, they are fine-tuned on KiSing for another 500 epochs. As discussed in Sec.~\ref{ssec: dataset}, Opencpop is 5.2-hour Mandarin pop singing voice database, which has the same language coverage with KiSing. As shown in Table~\ref{tab: transfer}, Multilingual$*$ achieves the best objective scores of MCD and F$_0$RMSE, while Opencpop$*$ achieves the best VUV\_E. For the subjective evaluation, we observe that both pre-trained models outperform the Baseline, while Multilingual$*$ is better than Opencpop$*$. The result indicates that larger data size and a larger coverage of phonemes could benefit the low-resource multi-style SVS. 

\begin{table}
\centering
\caption{\label{tab: transfer} Objective evaluation for transfer learning experiments on KiSing (a low-resource multi-style singing database). All the models use the same XiaoiceSing architecture. 
Models with $*$ are first pre-trained on the corresponding database and then fine-tuned on KiSing. 
}
\vspace{-10pt}
\begin{tabular}{l|cccc}
\toprule \textbf{Model} &\textbf{MCD$\downarrow$} & \textbf{VUV\_E$\downarrow$} & \textbf{F$_0$RMSE$\downarrow$}  \\ 
\midrule
Baseline   & 8.74 & 3.74 & 0.214 \\
Opencpop$*$  & 8.52 & \textbf{3.54} & 0.190\\
Multilingual$*$  & \textbf{8.25} & 3.66 & \textbf{0.177} \\
\bottomrule
\end{tabular}
    \vspace{-10pt}
\end{table}

\begin{figure}[tbp]
    \centering
    \centerline{\includegraphics[width=8.5cm]{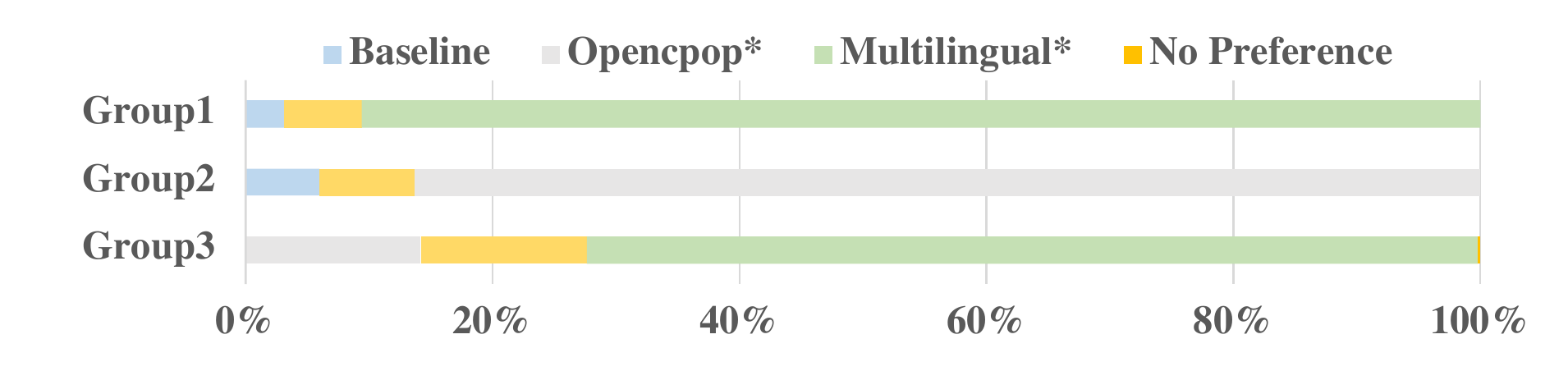}}
    \vspace{-10pt}
\caption{A/B test for transfer learning experiments on KiSing.}
\label{fig:abtest}
\vspace{-20pt}
\end{figure}




%% file: sections/4_Conclusion.tex
\vspace{-5pt}
\section{Conclusions}

In this work, we introduce a new E2E-SVS toolkit, called Muskits. This toolkit is developed as an open platform which can support fair comparison of different SVS models, and we expect it to facilitate the research field of SVS. The toolkit is based on the ESPnet framework with 12 recipes covering over 10 public databases in 4 languages. We conduct experiments under four scenarios: single-singer, multi-singer, multilingual SVS, and transfer learning on a multi-style database. In single- and multi-singer scenarios, we compare three published works on SVS with the same music score representation and vocoder. In addition, we explore multilingual SVS by combining six databases in four languages. From multilingual pre-training, we demonstrate subjective and objective gain on a single-singer database, which could alleviate the low-resource issue in SVS. Further experiments on transfer learning also show the effectiveness of our framework on a multi-style low-resource database. Full documentation and real-time demos can be found at \url{https://github.com/SJTMusicTeam/Muskits}.

\vspace{-5pt}
\section{Acknowledgement}

This work was supported by the National Natural Science Foundation of China (No. 62072462) and the National Key R\&D Program of China (No. 2020AAA0108600).